\documentclass{article}
\usepackage{a4}
\usepackage{amssymb}
\usepackage{graphicx}

\begin{document}
\title{Regularities in electromagnetic decay widths}
\author{Alejandro RIVERO \thanks{EUPT, 
Univ de Zaragoza, Campus de Teruel, 
 44003 Teruel, Spain} \thanks{email: arivero@unizar.es}}

\maketitle

\begin{abstract}
We revisit Sakurai's remark on the regularities of lepton-pair widths for
mesons, extending the panorama to radiative $X \to \gamma \gamma$ decays. 
The regularities persist, and somehow surprisingly
some of them seem to relate with Fermi's constant -or $Z^0$-.
\end{abstract}

Back in 1978, Sakurai took the opportunity of a festscrift  \cite{Sakurai:1978xb} to 
remark how the study of 
vector meson decays $V \to \gamma \to e^+ e^-$, when extended to $J/\Psi$ and $\Upsilon$,
seemed to have confirmed an empirical rule of Yennie \cite{Yennie:1974ga} about 
the universality of  $\Gamma (V \to e^+ e^-)$. On other hand a universality in the effective coupling 
$g_{X\gamma\gamma}$ for pseudoscalars $X=\pi,\eta$ is well known popularly and we have recently
noticed that it seems to extend to vector mesons, particularly to $J/\Psi$, when interpreted
as a virtual process for total decay width. And also radiative transitions as $\Sigma^0 \to
\Lambda^0 \gamma$ happen to be in the adequate range. 

Our purpose here is to review all the radiative decays well measured in the modern data tables
in order to establish how significant these regularities are.

\section{All total decay widths}
We can try to get some perspective by plotting all the decay rates of subnuclear 
particles. This can be done straightly from the tables provided by the particle data group \cite{pdg}
in its website {\tt http://pdg.lbl.gov/}. One can see three layers corresponding to decays strong, 
electromagnetic or weak, with the neutron being appart
from the rest of weak
decaying particles due to its quark composition.
The area of strong decay is actively pushed forward by experimentalists, 
and one can see how the actual advances hit on one hand the need of 
having a noticeable width in the resonance plot, 
say $\Gamma<E_0$, and on the other hand enough area. We have plotted 
the lines $E_0 \ \Gamma  = 
(1 GeV)^2$, $\Gamma= k E_0$ ($k=1, 0.5, 0.25$) as reference, but we 
excuse ourselves about
plotting the mass against spin dependence famously reported
in Chew-Frautschi plots.
 
When particles can not decay strongly -either because there are no other
hadronic mode available, or because any available mode
should change quark flavours or because asymptotic freedom activates
OZI rule- they give way to electromagnetic decay or to weak decay.  

In the weak decay area spectator models and other approaches (even plain dimensional analysis) point to a $\Gamma \sim m^5$ scaling, and we have drawn one such line from the muon. Note that such line in a spectator model should be used to meet the mass value of the decaying quark, not
of the whole particle. 

\begin {figure}
%\centering
\includegraphics[width=26cm]{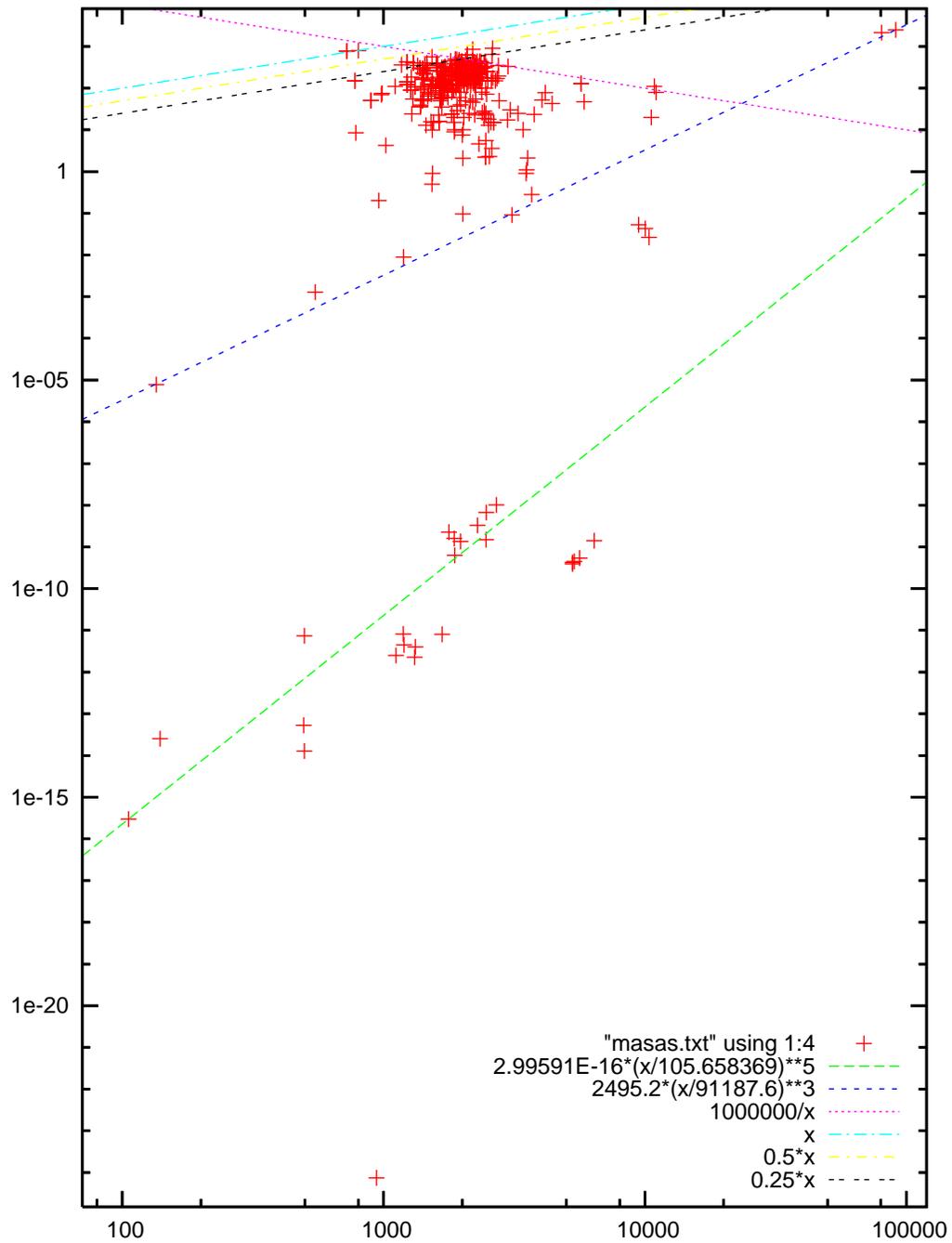}
\caption{Log-log plot of  $\Gamma(M)$ (units in MeV) from data in 
 {\tt http://pdg.lbl.gov/2005/mcdata/mass\_width\_2004.csv}.}
\label{fig1}
\end {figure}

\begin {figure}
%\centering
\includegraphics[width=12.5cm]{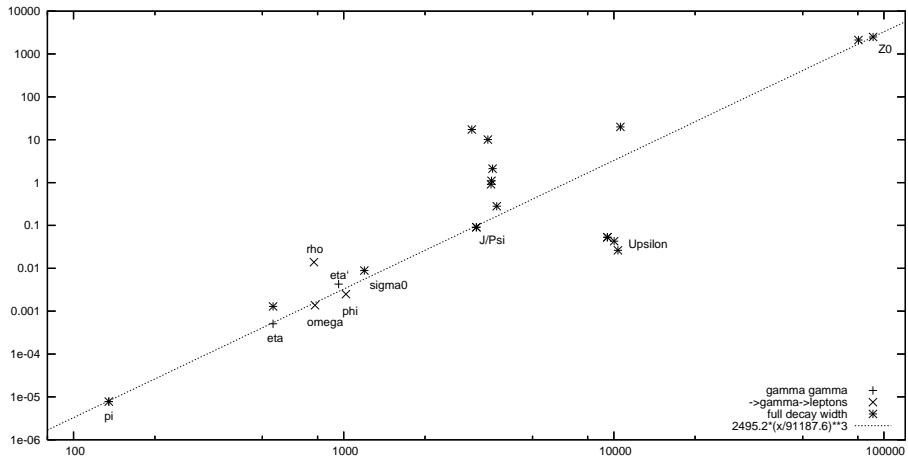}
\caption{Widths of electromagnetic decaying particles, including excited states of
charmonium and bottomonium, and its approximate cubic dependence of mass. Note that we have
included $\Upsilon(4S)$, which is strongly decaying, and also two points for $\eta$ (total decay
and $\gamma \gamma$ only).}
\label{fig2}
\end {figure}

The intermediate area, electromagnetic or radiative 
decays (we are using both terms as 
exchangeable ones in this context), is the one we are interested. We 
have marked it in both figures \ref{fig1}, \ref{fig2} with a
scaling $\Gamma \propto m^3$, partly because of the usual $\pi^0$ decay 
calculation,
partly because of dimensional comparison with weak decays: the weak decays
have gained a dimensionful coupling constant (Fermi constant) from
the breaking of electroweak symmetry, and then an additional $\propto m^2$.

It is interesting to stop a little to think how should the points in the
plot move if electroweak symmetry were restored, playing with the parameters in
the Higgs sector, thus altering the value of Fermi constant until a
phase transition is reached and it disappears. It is even possible to use 
mass formula of mesons and baryons to determine when a given strong
decay becomes available/unavailable and some particle leaves/enters the
electroweak zone.

Also it is a funny try to look for alternative visualizations of the non
strong sector (hint: in modern notation, the strong sector particles are the
ones that carry a mass value between parenthesis as part of its name). It has
been suggested \cite{MacGregor:2005tn} to use base 137 for the logs in the decay
width, and also\footnote{Amateur suggestion of Yuri Danoyan} to center the 
masses via an arctan(x/$M_0$) map, adjusting $M_0$ to the value of the proton.

\section{Non weak, non strong decays}

The list of known particles whose main decay mode is neither strong nor 
weak is very short. In order of increasing mass we meet:

-The neutral pion

-$\eta$, although it has already a mix with pions that makes use of the strong
force.

-$\Sigma^0$, the only baryon in the party.

-Lower charmonium, this is $J/\Psi$, $\Psi(2S)$, $\eta_c$ (mostly strong 
mixed), and some 1P wave states: $\chi_{c0},\chi_{c1},\chi_{c2}$.

-Lower bottomonium, this is $\Upsilon(1S)$, $\Upsilon(2S)$, $\Upsilon(3S)$

-$Z^0$ and $W^\pm $. Well, why not? They do not decay strongly, and we can
not say strictly that they decay weakly, no more than we can claim that
Death is dead.

If we calculate the reduced decay widths $\beta\equiv \Gamma/M^3$ for
the above group, only the $\Upsilon$ are noticeably appart.

\begin{table}
\begin{center}
\begin{tabular}{l|r}
particle  &  $\beta^{-\frac12}$ (GeV) \\
\hline       
$\pi^0$  &  561  \\ 
$\eta$  &  357 \\
$\Sigma^0$  & 138    \\
$J/\Psi$  &  571 \\
$\Psi(2S)$  & 422  \\
$\Upsilon(1S)$  &  3996  \\
$W^\pm$  &  495  \\
$Z^0 $  & 551  \\
\end{tabular}
\end{center}

\caption{[Inverse square root of] Reduced full decay widths $\beta=\Gamma/M^3$ for
 EM (generically, non strong non weak) decaying particles}
\end{table}

As I have noted in the introduction, and we will see now in the next
section, the coincidence between $\pi^0$ and $\eta$ increases if we only
consider $\gamma \gamma$ decay.

\section{All the non negligible radiative widths}

%en el plot de aqui, incluir la full decay width de todas y etiquetar
%cada ``columna''

\subsection{Decay via $\gamma$ to lepton pairs}

For electron pair production, this was reported by Sakurai \cite{Sakurai:1978xb}. Today
we can complete it with decays to muons and (for $\Upsilon$) tau.

%electron (+ muon?) pairs (faltan J/Psi y Upsilon)                        
\begin{tabular}{|c|r|r|l|l|}
particle & mass & total $\Gamma$   & $\Gamma (l^+l^-)$  &  $\Gamma (e^+e^-)$ \\   
\hline
$\rho$(770)  & 775.8  & 150.3     &            .01385766  &     .00701901 \\  %.0000467 + 0.0000455
$\omega(782)$  &  782.59 & 8.49       &       .001373682         & .000609582  \\ % .0000718 + 0.000090   
$\phi$(1020)   &  1019.46 &4.26       &            .00248784     & .00126948\\ %.000298 + 0.000286 
J/$\Psi$  & 3097   & 0.091     &             .0107471       &.0053963 \\ %.0593 + 0.0588    
$\Upsilon$ (1S)  &  9460   & 0.053   &       .0039909    &   .0012614   \\   %0.0238 +0.0248 +0.0267   
\end{tabular}

The analysis of aimed to conclude that $\Gamma$ was almost constant in this kind of decays. We are
more interested in the fact that the amplitudes have values near the cubic scaling. This is even
truer if following \cite{Yennie:1974ga} we consider a proportion 1:9 between $\rho$ and $\omega$ 

\subsection{Decay to $\gamma \gamma$}

This kind of decay -- and optionally its descendant $\gamma e^+ e^-$ -- is useful to separate strong force
effects from electromagnetic ones in decays of $\eta$ and $\eta'$; see \cite{Feldmann:1999uf} for 
calculations and estimates of mixing effects. It has also been used to argument against a quark 
composition of scalars $a_0$, $f_0$, on the basis of a failure of $m^3$ scaling \cite{Kalashnikova:2005zz}. We
omit this pair of scalars because their decay and total width is not completely stablished yet. 

Besides, it does exist data about
this kind of decay for $\eta_c$ and $\eta_c(2S)$ and for some spin 2 particles $a_2$, $f_2$, $f'_2$ which we 
list for completion.

\begin{center}
\begin{tabular}{|c|r|l|}
particle &  mass & $\Gamma(\gamma\gamma)$ \\  %
\hline 
$\pi^0$ &134.97    &  .00000778 \\ %         .0000078  (.99798 + .0198)->.000007938 (or.00000778)->->    .000001796 or .000001779
$\eta$ &547.75    &  .000509   \\  %     0.00129 *(.3943+.0060)  -> .000516 (or .000509) ->     -> .00000177 (+muons faltan?)
$\eta'$&957.78    &   .0042824 \\  %          .202 * .0212             -> .0042824             ->     -> .00000221
$f_2$ &1275.4     & .00260991  \\  %          185.1 .0000141            -> .00260991            ->     -> .00000112
$a_2$  &  1318.3   &  .0010058 \\  %           107       .0000094     -> .0010058             ->     -> .0000006625
$f'_2$ & 1525      &  .00008103 \\ %                73\pm5        .00000111    -> .00008103        ->     -> .000000151153
$\eta'_c$ & 3642.9  &      .0013 \\ %              .        -> .0013       ->     -> .0000001639
$\eta c$  &      2980   & .0074 \\ %                   >pdg          -> .0074              ->  .00000052
\end{tabular}
\end{center}

\subsection{Decay to anything plus $\gamma$}
 
The particle data group list decays to $X\gamma$ for about a dozen of strongly decaying mesons but
the corresponding widths are still scattered along various orders of magnitude and some extra organizing
principle is still needed.

%rho+    775.8   150.3                 .00045 
%rho770  775.8   150.3                 .0108
%w782    782.59  8.49                  .0892
%eta'    957.78  .202                  .325        
%phi     1019.46 4.26                  .013
%b1      1229.5  142                   .0016 
%f1      1281.8  24.1                  .055
%a2      1318.3  107                   .00268       
%K*      896.10  50.7                  .00230
%K*pm    891.66  50.8                  .00099
%K*2pm   1425.6  98.5                  .0024
%D*      2010.0  .096                  .016
%l1520   1519.5  15.6                  .008
%s0      1192.6   ///                  .999

For the baryons, the dominant (100\%) $\Sigma^0 \to \gamma \Lambda^0$ is the only decay
measured directly, although
via Primakoff method, and it fits very nicely in the expected range. We could try to add some
 decays built from the fit of $N$ and $\Delta$ resonances,
for instance from the database of \cite{Arndt:2002xv} (see \cite{Capstick:1992uc} for a theoretical
model and
some experimental plots too), but we felt so blocked as in the meson case.

\subsection{Whole widths}

Besides the above cases of $\Sigma^0$ and $\pi^0$ whose radiative decay is practically the whole decay, we
are interested on considering the Charmonium and Bottomonium areas, where OZI rule (a combination of
asymptotic freedom and conservation laws) forbids strong decay and hadron production also proceeds 
mainly via the electroweak forces.
We find the amazing fact that $J/\Psi$ total width scales respective to $Z^0$ (!) total decay as $m^3$, while
$\Upsilon$ could be thought to scale from it as $m^5$ (but the states of $\Upsilon$ having allowed strong
decays are closer to the cubic scaling than to quintic one).

\section{Discussion}

The plot in figure 2 summarizes the observations --we can barely call them results. 

While the decay to 
leptons stated by Sakurai persists, it is to us
be of a minor consideration when one takes into account that the whole $J/\Psi$ decay must be electromagnetically
dominated. In this case the property that becomes interesting is the alignment according a scaling $\Gamma \propto M^3$
already noticed in \cite{Rivero:2005ky} and which now incorporates Sakurai' vector meson decays for $\omega$ and
$\phi$ as well as the baryon $\Sigma^0$. Besides, the consideration of the decay to $\gamma \gamma$ of $\eta$ and
$\eta'$, instead of the total width, lets one to put such particles near the approximate scaling rule, but this
was perhaps already known \cite[pg 10]{Shore:2006mm} in the decay lore. In total we have eight electromagnetically decaying strong 
particles (ten if we count $W^+$ and
$\Psi(2S)$ following a scaling rule across four orders of magnitude in mass and mysteriously fitting a purely electroweak
quantity, the decay of $Z^0$ (that is controlled by $\sin \theta_W$ basically\cite{Rivero:2005bq}). Among spin 0 and spin 1 mesons,
only $\Upsilon$, $\rho$ and some excited states do not fit straightly into this rule, and really kinematic and symmetry arguments
could be invoked to fit every except $\Upsilon$; note for instance already in \cite{Yennie:1974ga} the expected 9:1 
proportion between $\rho$ and $\omega$.

As a collateral remark, it is noticeable that the double gamma decay of $\eta_c$ resolves near of the leptonic
decay of $J/\Psi$; perhaps a rule can be built relating both.

I want to thank K. Illinsky from the PDG by providing a corrected version of the file {\tt mass\_width\_2004.csv}; this
work was retaken as a way to double check the values of such file.


\begin{thebibliography}{20}

%\cite{Arndt:2002xv}
\bibitem{Arndt:2002xv}
  R.~A.~Arndt, W.~J.~Briscoe, I.~I.~Strakovsky and R.~L.~Workman,
  %``Analysis of pion photoproduction data,''
  Phys.\ Rev.\ C {\bf 66}, 055213 (2002)
  [arXiv:nucl-th/0205067].
  %%CITATION = NUCL-TH 0205067;%%

%\cite{Asner:2003wv}
\bibitem{Asner:2003wv}
  D.~M.~Asner {\it et al.}  [CLEO Collaboration],
  %``Observation of eta/c' production in gamma gamma fusion at CLEO,''
  Phys.\ Rev.\ Lett.\  {\bf 92} (2004) 142001
  [arXiv:hep-ex/0312058].
  %%CITATION = HEP-EX 0312058;%%

%\cite{Brown:1988ms}
\bibitem{Brown:1988ms}
  R.~M.~Brown,
  %``Weak Decays Of Stable Particles,''
  Rept.\ Prog.\ Phys.\  {\bf 51}, 1373 (1988).
  %%CITATION = RPPHA,51,1373;%%

%\cite{Brown:1987rn}
%\bibitem{Brown:1987rn}
%  R.~W.~Brown and E.~A.~Paschos,
%``Relation Between The Radiative And Pionic Weak Decays Of Hyperons,''
%  Nucl.\ Phys.\ B {\bf 319}, 623 (1989).
%%CITATION = NUPHA,B319,623;%%

%\cite{Capstick:1992uc}
\bibitem{Capstick:1992uc}
  S.~Capstick,
  %``Photoproduction and electroproduction of nonstrange baryon resonances in
  %the relativized quark model,''
  Phys.\ Rev.\ D {\bf 46} (1992) 2864.
  %%CITATION = PHRVA,D46,2864;%%

%\cite{Feldmann:1999uf}
\bibitem{Feldmann:1999uf}
  T.~Feldmann,
  %``Quark structure of pseudoscalar mesons,''
  Int.\ J.\ Mod.\ Phys.\ A {\bf 15} (2000) 159
  [arXiv:hep-ph/9907491].
  %%CITATION = HEP-PH 9907491;%%

%\cite{Gounaris:1976wp}
\bibitem{Gounaris:1976wp}
  G.~J.~Gounaris,
  %``SU(3) Invariance In Radiative Decays And The Vector Meson Masses,''
  Phys.\ Lett.\ B {\bf 63}, 307 (1976).
  %%CITATION = PHLTA,B63,307;%%

%\cite{Kalashnikova:2005zz}
\bibitem{Kalashnikova:2005zz}
  Y.~Kalashnikova, A.~Kudryavtsev, A.~V.~Nefediev, J.~Haidenbauer and C.~Hanhart,
  %``Insights on scalar mesons from their radiative decays,''
  arXiv:nucl-th/0512028.
  %%CITATION = NUCL-TH 0512028;%%

%\cite{MacGregor:2005tn}
\bibitem{MacGregor:2005tn}
  M.~H.~Mac Gregor,
  %``Electron generation of leptons and hadrons with reciprocal alpha-quantized
  %lifetimes and masses,''
  Int.\ J.\ Mod.\ Phys.\ A {\bf 20} (2005) 719
  [arXiv:hep-ph/0506033].
  %%CITATION = HEP-PH 0506033;%%

%\cite{Rivero:2005bq}
\bibitem{Rivero:2005bq}
  A.~Rivero,
  %``GUT angle minimises Z0 decay,''
  arXiv:hep-ph/0511165.
  %%CITATION = HEP-PH 0511165;%%

%\cite{Rivero:2005ky}
\bibitem{Rivero:2005ky}
  A.~Rivero,
  %``Anomaly-driven decay of massive vector bosons,''
  arXiv:hep-ph/0507144.
  %%CITATION = HEP-PH 0507144;%%

%\cite{Sakurai:1978xb}
\bibitem{Sakurai:1978xb}
  J.~J.~Sakurai,
  %``Remarkable Regularity In The Lepton - Pair Widths Of Vector Mesons,''
UCLA/78/TEP/20
%\href{http://www.slac.stanford.edu/spires/find/hep/www?r=ucla\%2F78\%2Ftep\%2F20}{SPIRES entry}

%\cite{Shore:2006mm}
\bibitem{Shore:2006mm}
  G.~M.~Shore,
  %``Pseudoscalar meson decay constants and couplings, the Witten-Veneziano
  %formula beyond large N(c,) and the topological susceptibility,''
  arXiv:hep-ph/0601051.
  %%CITATION = HEP-PH 0601051;%%

%\cite{Singh:1981pg}
\bibitem{Singh:1981pg}
  C.~P.~Singh and P.~K.~Chatley,
  %``Radiative Decays Of Vector Mesons,''
  Phys.\ Rev.\ D {\bf 26}, 332 (1982).
  %%CITATION = PHRVA,D26,332;%%

%\cite{Yennie:1974ga}
\bibitem{Yennie:1974ga}
  D.~R.~Yennie,
  %``Comment On Radiative Corrections To E+ E- $\to$ Psi (3105),''
  Phys.\ Rev.\ Lett.\  {\bf 34}, 239 (1975).
  %%CITATION = PRLTA,34,239;%%

\bibitem{pdg}
 S. Eidelman et al., Physics Letters B592, 1 (2004)
and 2005 partial update for edition 2006



\end{thebibliography}
\end{document}